\documentclass[aps,prl,twocolumn,showpacs,groupedaddress]{revtex4}
\usepackage{graphicx}
\usepackage{dcolumn}
\usepackage{bm}
\usepackage{psfig,graphics}
\usepackage{latexsym}
\usepackage{amsmath}
\usepackage{amssymb}
\usepackage{enumerate}
\usepackage[latin1]{inputenc}
\usepackage{bbm}

\newcommand{\rot}{{\cal R}}

\newcommand{\bra}[1]{\ensuremath{\langle#1|}}

\newcommand{\ket}[1]{\ensuremath{|#1\rangle}}

\newcommand{\kommentar}[1]{}

\newcommand{\mean}[1]{\ensuremath{{\langle #1 \rangle}}}

\newcommand{\be}{\begin{equation}}
\newcommand{\ee}{\end{equation}}
\newcommand{\ben}{\begin{displaymath}}
\newcommand{\een}{\end{displaymath}}
\newcommand{\bea}{\begin{eqnarray}}
\newcommand{\eea}{\end{eqnarray}}
\newcommand{\bean}{\begin{eqnarray*}}
\newcommand{\eean}{\end{eqnarray*}}
\newcommand{\bc}{\begin{center}}
\newcommand{\ec}{\end{center}}
\newcommand{\bi}{\begin{itemize}}
\newcommand{\ei}{\end{itemize}}
\newcommand{\vecb}[1]{{\bf #1}}
\newcommand{\cf}{cf.}

\begin{document}

\title{Differential atom interferometry beyond the standard quantum limit}

\author{K. Eckert\mbox{$^{1}$}, P. Hyllus\mbox{$^{1,2}$},
D. Bru{\ss}\mbox{$^{1,3}$}, U.V. Poulsen\mbox{$^{1,4,5}$}, 
M.~Lewenstein\mbox{$^{1,6,}$}\footnote{also at Instituci\'o
Catalana de recerca i estudis avan\c cats.}, 
C. Jentsch\mbox{$^{7}$}, T.
M{\"u}ller\mbox{$^{7}$}, E.M. Rasel\mbox{$^{7}$}, and W.
Ertmer\mbox{$^{7}$}} 
\affiliation{ \mbox{$^{1}$}Institut f{\"u}r Theoretische Physik, Universit\"at Hannover, D-30167 Hannover, Germany.\\
\mbox{$^{2}$} The Blackett Lab -- QOLS, Imperial College London, London SW7 2BW, United Kingdom.\\
\mbox{$^{3}$}Institut f{\"u}r Theoretische Physik III, Universit{\"a}t D{\"u}sseldorf, D-40225 D{\"u}sseldorf, Germany.\\
\mbox{$^{4}$}Dipartimento di Fisica, Universit{\`a} di Trento, I-38050 Povo (TN), Italy;
and ECT$^*$, I-38050 Villazzano (TN), Italy.\\
\mbox{$^{5}$}Department of Physics and Astronomy, University of Aarhus, DK-8000 Aarhus C., Denmark.\\
\mbox{$^{6}$}ICFO - Institut de Ci\`encies Fot\`oniques, 08034 Barcelona, Spain.\\
\mbox{$^{7}$}Institut f\"ur Quantenoptik, Universit{\"a}t Hannover, D-30167 Hannover, Germany. }

\date{\today}

\begin{abstract}
We analyze methods to go beyond the standard quantum limit for a class
of atomic interferometers, where the quantity of interest is the
difference of phase shifts obtained by two independent atomic
ensembles. An example is given by an atomic Sagnac interferometer,
where for two ensembles propagating in opposite directions in the
interferometer this phase difference encodes the angular velocity of the
experimental setup.
We discuss methods of squeezing separately or jointly
observables of the two atomic ensembles, and compare in detail
advantages and drawbacks of such schemes. In particular we show that the method of
joint squeezing may improve the variance by up to a factor of 2. We take into account
fluctuations of the number of atoms in both the preparation and the
measurement stage, and obtain bounds on the difference of
the numbers of atoms in the two ensembles, as well as on the detection
efficiency, which have to be fulfilled in order 
to surpass the standard quantum limit. 
Under realistic conditions,
the performance of both schemes can be improved
significantly by reading out the phase difference {\em via} a 
quantum non-demolition (QND) measurement.
Finally, we discuss a scheme using macroscopically entangled
ensembles.

\end{abstract}


\maketitle
\date{\today}
\newpage

\section{I. Introduction}

Comparing the phase shifts obtained in two independent interferometric
setups has several applications, prominent examples being the comparison of
atomic clocks \cite{atomicclocks}, and Sagnac interferometry to discriminate
between rotations and accelerations \cite{jentsch}. In addition, the 
comparison of gravitational forces at different points in space or for
different atomic species \cite{weitz} allows to test predictions of possible
violations of Einstein's general relativity \cite{snadden:1998,kasevich:2002}.
For such {\em differential interferometers}, the
quantity of interest is encoded in either the difference, or the sum
of the individual phase shifts. 

Especially for the measurement of inertial forces, atom interferometers
promise high resolution. Here atom optical elements like beam splitters
and mirrors can be realized using Raman transitions between two
atomic ground state levels \cite{KasevichChu}.
The accumulated phase difference
between the interferometer paths is encoded in the number
difference of atoms in the exit ports labeled by different
internal states. For example, this can be measured by state
selective fluorescence detection. Such schemes have already been
successfully implemented to measure inertial forces and the
earth's gravity with a high accuracy~\cite{KasevichChu,peters}. In
Sagnac atom interferometry, the goal is to measure the phase shift
which occurs when the laser setup (laboratory frame) is rotating
relative to the frame of the freely flying atomic ensembles. This phase shift is given by
$\phi^{\rm at} =4\pi{\bf A}\;\mbox{\boldmath $\Omega$}\;m_{\rm
at}/h$ as compared to $\phi^{\rm light}=4\pi{\bf A}\;
\mbox{\boldmath $\Omega$}/(\lambda c)$ for laser interferometers,
where $m_{\rm at}$ is the mass of the atoms, ${\bf A}$ is the
oriented enclosed area of the interferometer, \mbox{\boldmath
$\Omega$} is the vector of angular velocity, and $\lambda$ is the wavelength of
the light. For an atomic gyroscope working with $^{87}$Rb, the phase
$\phi^{\rm at}$ is $10^{11}$ times larger than the corresponding phase
$\phi^{\rm light}$ of a light interferometer enclosing the
same area and operating at $\lambda=10^3\;$nm. Hence, atom
interferometers promise an enormously improved resolution for rotation
measurements as compared to ``classical'' photonic devices.

The Sagnac phase can be measured by letting two
ensembles of atoms pass through the interferometer from
opposite sides. They obtain then phase shifts $\pm\phi$
due to the rotation of the laboratory frame, where the sign depends
on the direction of propagation of the ensembles, and 
a common phase shift $\theta$ due to effects such as 
an acceleration of the setup. 
Subtracting the phases of the two ensembles yields the desired phase
$2\phi$, which encodes the rotation of the setup around an axis perpendicular
to the plane of the interferometer. 
Collisions between the two ensembles in the
interferometer can safely be neglected due
to the low atomic densities of the ensembles.

In the standard quantum limit for phase measurements in atomic
interference experiments, the variance of the phase due
to quantum projection noise is given by 
$(\Delta\phi)^2_{\rm SQL}=1/N,$ where $N$ is the number of 
atoms in a sample. By
feeding the interferometer with non-classical states of atoms, the
so-called squeezed states, this limit can be surpassed,
with a fundamental bound given by
$(\Delta\phi)^2_{\rm H}=1/N^2,$ the so-called
Heisenberg limit \cite{inter,heisenberg,oblak}. 
Squeezed atomic states can be produced by a quantum non-demolition
(QND) interaction of the atoms with a light beam \cite{QND}, or by 
absorption of non-classical
states of the light \cite{absorbnonclassical}. 
Squeezed states of atomic ensembles might also be useful as
quantum memory for states of light \cite{memory}. 

It has been shown that for such squeezed states atoms within an ensemble are
entangled with each other \cite{Soerensen}. In addition to this
entanglement on a microscopic level, it is also possible to
entangle macroscopic degrees of freedom of two atomic ensembles
with a similar interaction \cite{Duan1,Polzik}. This in principle
enables teleportation of the macroscopic state of an ensemble.
Furthermore, it has been shown recently that such macroscopically
entangled ensembles can improve the efficiency of measurements 
of the components of a magnetic field \cite{Vivi}. 
Here, we try to exploit microscopic as well as macroscopic 
entanglement between two atomic ensembles in the context 
of differential interferometry, where we focus especially on an 
atomic Sagnac interferometer setup.

In Section II we calculate the phase variance for a differential
interferometer using non-squeezed coherent states in order to
introduce our methodology. We show also how to include number
fluctuations into the calculations. These will turn out to be important later.
In Section III we show that the phase uncertainty can be reduced by
feeding the interferometer with individually squeezed ensembles. In
Section IV we consider squeezing of a joint observable of both ensembles.
In Section V we discuss how decoherence affects the interferometer,
and in Section VI we compare the variances of the
schemes discussed so far for realistic parameters. Finally, in
Section VII the use of macroscopically entangled states in the
interferometer is discussed.

\section{II. Coherent input states}
For a single atom we define a pseudo spin-$1/2$ through two ground state
atomic hyperfine levels $\ket{1}$ and $\ket{2}$ \cite{jentsch,kasevich:2002} 
by introducing the relevant spin operators as
\begin{eqnarray}\label{eqs:defsigma}
\hat{\sigma}_z&=&\frac{1}{2}\left(\ket{1}\bra{1}-\ket{2}\bra{2}\right)\;,
\quad
\hat{\sigma}_x=\frac{1}{2}\left(\ket{1}\bra{2}+\ket{2}\bra{1}\right)\:,
\nonumber\\
\hat{\sigma}_y&=&\frac{1}{2i}\left(\ket{1}\bra{2}-\ket{2}\bra{1}\right).
\end{eqnarray}
We will subsequently only consider the collective spin
$\hat{\bf J}=\sum_{i=1}^{N_J}\hat{\mbox{\boldmath $\sigma$}}{\,(i)}$
for the first, and in analogy $\hat{\bf L}$ for the second ensemble;
$N_{J,L}$ are the number of atoms in the ensembles $J$ and $L$, respectively.
The measurements that can typically be performed
in atom interferometry are population measurements
of the levels $\ket{1}$ and $\ket{2}$ by fluorescence
techniques \cite{Kasevich}. Hence, only the $z$
components of the collective spin vectors can be
measured directly.

\subsection{II.A. Description of the interferometer}

Initially, all the atoms are assumed to be prepared in the
state $\ket{1}$, leading to the following expectation
value and variance of the collective spin vector:
\bea
    \mean{\hat{\bf J}}=\frac{N_J}{2}\hat{\bf z},\quad
    (\Delta \hat{J}_z)^2=0,\quad (\Delta \hat{J}_{x,y})^2=\frac{N_J}{4},
    \label{eqs:Jin}
\eea
Analogous values are obtained for the ensemble $\hat{\bf L}$.
Corresponding to the
definitions in Eq.~(\ref{eqs:defsigma}), the $z$ axis denotes the
difference of the number of atoms in states $\ket{1}$ and
$\ket{2}$, whereas the phase difference between these two states
is encoded in the $x-y$ plane. The uncertainties stem from
the single particle uncertainties, and correspond to a state with a {\em fixed}
number of atoms.
%
%

\begin{figure}[th]
\includegraphics[width=6.5cm]{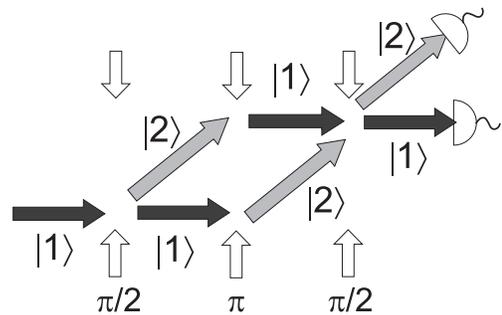}
\caption{Scheme of the atom interferometer. $\frac{\pi}2$ and $\pi$
label the beamsplitting and the mirror pulses, respectively. The populations
in the two exit ports are detected state-selectively {\it via} fluorescence
measurements.
}
\label{fig:AI}
\end{figure}

A typical interferometer sequence used for the measurement of
inertial forces consists of three atom--light interactions as
shown in Fig.~\ref{fig:AI}. The first beam splitting 
Raman pulse transfers all the atoms from the ground state
$\ket{1}$ to the superposition
$\frac{1}{\sqrt{2}}(\ket{1}+\ket{2})$. Atoms transfered to state
$\ket{2}$ obtain a momentum kick of two photon recoil
if the two Raman lasers are counter-propagating, so
that the partial waves delocalize, as depicted in Fig.~\ref{fig:AI}. The
second pulse exchanges the populations,
$\ket{1}\leftrightarrow\ket{2}$, and deflects the partial waves.
Finally, they are recombined  in the last interaction zone, acting as a
beam splitter. 
%
These pulses can be represented as rotations of the collective
spin vectors $\mean{\hat{\bf J}}$ and $\mean{\hat{\bf L}}$ around an axis in the $x-y$
plane, the angle being given by $\pi/2$ for beam splitters and by
$\pi$ for mirrors. For a fixed coordinate system the angle between
the $x$ axis and the rotation axis is given by the laser phase
\cite{atominterferometry}. The laser phases change if the setup (laboratory frame)
rotates with respect to the path of the freely flying ensembles,
which causes the Sagnac phase shift. This
change corresponds to a rotation of the collective spin vectors in
the $x-y$ plane around the $z$ direction.

In order to make the scheme applicable to a more general scenario for
differential interferometers, we model it as follows for each of the
ensembles labeled by $\hat{\bf J}$ and $\hat{\bf L}$, respectively:
the first beam splitter rotates each collective spin
vector by $\pi/2$ around the $y$ axis, then we collect all the
phase shifts occurring in the interferometer in rotations around $z$
of $\hat{\vecb{J}}$ and $\hat{\vecb{L}}$ by $\Phi_J$ and $\Phi_L$,
respectively. Finally, a $\pi$ and a $\pi/2$ pulse, both 
around $x$, implement the mirror and the final
beam splitter, respectively. These last two pulses can
be combined to a single rotation around the $x$ axis by $-\pi/2$.
\begin{figure}[h]
\includegraphics[width=0.9\columnwidth]{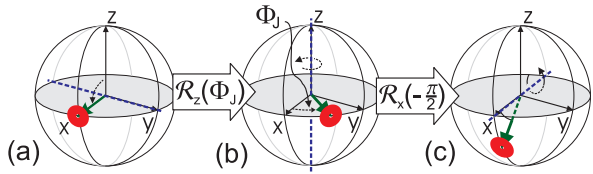}
\caption{(Color online) Interferometric scheme for a coherent input state:
(a) collective spin after the first beam splitter, (b) the total phase $\phi$
accumulated in the interferometer results in a rotation around $z$,
before (c) a final rotation around $x$ by $-\pi/2$ implements the 
mirror and the final beam splitter, encoding the phase in the 
$z$ direction.}
\label{fig:CS}
\end{figure}
All relevant interferometric steps are described in Fig.~\ref{fig:CS},
where the atomic spin vector is depicted together with a disc 
representing the corresponding uncertainties of the spin operators.
To simplify notation we will take the state after
the first beam splitter as the initial state:
$\hat{\vecb{J}}^{\rm in}=\rot_y(\pi/2)\hat{\vecb{J}}$, see Fig. \ref{fig:CS}(a).
Here $\rot_{i}(\alpha)$, $i\in\{x,y,z\}$, is the matrix rotating a vector
by the angle $\alpha$ around the direction $\hat{\bf e}_i$. 
In the Heisenberg picture, and neglecting collisions between the atoms
within the ensemble, the spin operator changes according to 
$\hat{{\bf J}}^{\rm out}=\rot_x(-\pi/2)\rot_z(\Phi_J)\hat{{\bf J}}^{\rm in}$.
This leads to
\be
    \hat{{\bf J}}^{\rm out}=\left(
    \begin{array}{c}
        \hat{J}_x^{\,\rm in}\cos\Phi_J -\hat{J}_y^{\,\rm in}\sin\Phi_J \\
        \hat{J}_z^{\rm in}\\
        -\hat{J}_x^{\,\rm in}\sin\Phi_J -\hat{J}_y^{\,\rm in}\cos\Phi_J 
    \end{array}
    \right).
\ee
It is also possible to consider a balanced atom interferometer
in which all rotations are around the $x$ axis. In this
case, an extra $\pi/2$ shift around $z$ leads to the same result.
The advantage of the extra pulse is that for the initial state of
Eq.~(\ref{eqs:Jin})
\be
    \mean{\hat{J}_z^{\rm out}}=-\frac{N_J}{2}\sin\Phi_J\approx-\frac{N_J}{2}\Phi_J
\ee
for small angles $\Phi_J$, while with the unmodified balanced
scheme $\mean{\hat{J}_z^{\rm out}}$ would be proportional
to $\cos\Phi_J$, and hence sensitive to $\Phi_J$ 
only in second order around $\Phi_J=0$.

As explained in the introduction, the phase
will be given by $\Phi_J=\phi+\theta$ for $\hat{\bf J}$
and by $\Phi_L=-\phi+\theta$ for $\hat{\bf L}$, because the
Sagnac phase $\phi$ takes a different sign, depending on
direction in which the ensemble passes the interferometer.
We define the phase operator for the
case of coherent states (cs) as
\be
    \hat{\phi}_{\rm cs}=-\frac{\hat{J}_z^{\rm out}}{N_J}+\frac{\hat{L}_z^{\rm out}}{N_L}
    \label{eqs:phiCS}
\ee
and obtain to the first order in $\phi$ and $\theta$
\bea
    \mean{\hat{\phi}_{\rm cs}}&=&\phi \\
    (\Delta\hat{\phi}_{\rm cs})^2&=&\frac{1}{4N_J}+\frac{1}{4N_L}.
\eea
The variance of $\hat{\phi}_{\rm cs}$ corresponds to
the standard quantum limit. Note that
$\theta$ can be obtained in an analogous way.
From the definition it is obvious that determining the
number of atoms is important for the calculation of the
phase shift. A major source of error will come from fluctuations
in the number of atoms of the ensembles and hence we will discuss
how to include this process into the calculations in the next section.

\subsection{II.B. Number fluctuations}
There are two sources of deviations of the number of atoms:
the preparation process and the number measurement process. The atomic
ensembles produced from the source are best described as a statistical mixture
of states with different atom numbers, but we will assume that the 
final number measurement projects
onto a number state with $N_J$ and $N_L$ atoms in the two
ensembles. Defining $\bar{N}=(N_J+N_L)/2$, we assume that
$|N_J-N_L|=\gamma\sqrt{\bar{N}}$, which reflects the variances of
the number operators $\hat{N}_{J,L}$ of the ensembles after the
production. Thus, $\gamma$ is the parameter which describes how 
well the atom numbers in the two ensembles match.

We treat the number measurements by introducing operators
$\delta \hat{N}$ with
\be
    \mean{\delta \hat{N}}=0,\quad [\Delta(\delta \hat{N})]^2=\alpha N,
\ee where $\alpha$ describes the quality of the number
measurement. For fluorescence measurements, $\alpha^{-1}$ is given
by the mean number of times an atom goes through the fluorescence
cycle and scatters a photon which subsequently is registered in the detectors
\cite{numbers}. Typical values in nowadays experiments 
are $\alpha^{-1}\approx 50\ldots100$. We replace $N_J$ by
$N_J^0+\delta \hat{N}_{J}^{(1)}+\delta \hat{N}_{J}^{(2)}$ and $\hat{J}_z^{\rm out}$ by
$\hat{J}_z^{{\rm out},0}+(\delta \hat{N}_{J}^{(1)}-\delta \hat{N}_{J}^{(2)})/2$. $N_J^0$
refers to the actual number that would have been
measured in perfect number projection measurements, and in $\delta\hat{N}_J^{(i)}$
the index $i$ corresponds to the atomic levels $\ket{i}$.

With these substitutions, we obtain
\bea
    \mean{\hat{\phi}_{\rm cs}}&=&\phi
        -\alpha\Big(\frac{1}{N_J^0}+\frac{1}{N_L^0}\Big)\\
    (\Delta\hat{\phi}_{\rm cs})^2&=&\frac{1}{4N_J^0}+\frac{1}{4N_L^0}
        +\alpha\Big(\frac{1}{N_J^0}+\frac{1}{N_L^0}\Big),
\eea
where terms of higher order have been neglected.
The contribution from the number fluctuations is in
agreement with Ref.~\cite{numbers}.
The expectation value of $\hat{\phi}_{\rm cs}$
is shifted for $\alpha\neq 0$. As this shift is 
of the order of the second term in 
$(\Delta\hat{\phi}_{\rm cs})^2$, it is of second order 
of the standard deviation only, 
and can thus safely be neglected. 
From the expressions it is clear that $\alpha\ll 1$ is 
required in order to reach the fundamental limit 
for the phase resolution using coherent input states.

In the following, we will drop the superscript $0$ on the
atom number, as long as there is no danger of confusion.

\section{III. Separately squeezed ensembles}

It is known that by taking squeezed input states
it is possible to surpass the standard quantum limit
in interferometry \cite{inter}. In this section, we
consider the case where both ensembles are squeezed
separately with the method introduced in \cite{QND},
{\it i.e.}, by a QND interaction with a laser beam shortly
after the first beam splitter.

\subsection{III.A. Squeezing a single ensemble}

\begin{figure}[h]
\includegraphics[width=4.5cm]{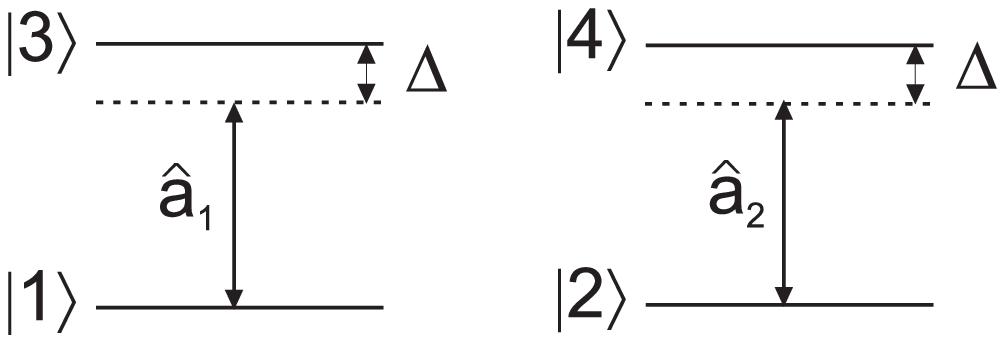}
\caption{For the QND interaction, the levels $\ket{1}$ and $\ket{2}$
	are coupled off-resonantly to states $\ket{3}$ and $\ket{4}$.}
\label{levels}
\end{figure}

We assume to have the situation of Fig.~\ref{levels},
{\em i.e.}, the electromagnetic field mode $a_{1}$
couples states $\ket{1}$ and $\ket{3}$,
and $a_{2}$
couples states $\ket{2}$ and $\ket{4}$.
Here transitions $\ket{1}\leftrightarrow\ket{4}$
and $\ket{2}\leftrightarrow\ket{3}$
have to be suppressed to the first order in the
coupling constant (\cite{QND}, see also the discussion
in Section V).
For the light, an effective spin vector can be defined, the so-called
Stokes vector. Its components are given by
\begin{eqnarray}
\hat{S}_z&=&\frac{1}{2}\left(\hat{a}_1^{\dagger}\hat{a}_1-\hat{a}_2^{\dagger}\hat{a}_2\right),
\ 
\hat{S}_x=\frac{1}{2}\left(\hat{a}_1^{\dagger}\hat{a}_2+\hat{a}_2^{\dagger}\hat{a}_1\right),
\\
\hat{S}_y&=&\frac{1}{2i}\left(\hat{a}_1^{\dagger}\hat{a}_2-\hat{a}_2^{\dagger}\hat{a}_1\right),
\end{eqnarray}
where $\hat{a}_1^{\dagger}$ and $\hat{a}_2^{\dagger}$ create a photon in mode $a_1$ and $a_2$,
respectively.
$\hat{S}_z$ measures the difference of photons in the two modes.
By an appropriate choice of the parameters, the
Hamiltonian describing the interaction of the light with the
first atomic
ensemble can be brought to the form \cite{QND}
\begin{equation}
  \hat{H}=\hbar\Omega_J \hat{S}_z \hat{J}_z,
  \label{HQND}
\end{equation}
with a frequency $\Omega_J$.
Due to the interaction, the collective spin vector of the
atoms is rotated around the $z$ axis by $\chi_J \hat{S}_z$,
with the atom--photon coupling $\chi_J=\Omega_J t$ and the effective interaction
time $t$.
The Stokes vector undergoes the same variation with
$\chi_J \hat{S}_z$ replaced by $\chi_J \hat{J}_z$. This 
rotation of the Stokes vector is due to the Faraday effect of the light 
passing the atoms, while the rotation of the atomic spin vector is due
to an AC Stark shift originating from the light field. The coupling
$\chi_J$ is given by \cite{QND}
\be
\chi_J=2 g^2\frac{L}{c}\frac{\Delta}{\frac14\Gamma^2+\Delta^2},
\quad g=\sqrt{\frac{\omega d^2}{2\hbar\epsilon_0 A L}},
\label{eq:chi}
\ee
where $A$ and $L$ are cross section and length of the atomic
sample, $\Gamma$ and $d$ are line-width and dipole moment 
of the atomic transition, respectively,
and $\Delta$ is the detuning from
the atomic resonance frequency $\omega$.

After the interaction,
\be
    \hat{S}_y^{\rm out}=\sin(\chi_J \hat{J}_z)\hat{S}_x^{\rm in}+\cos(\chi_J \hat{J}_z)\hat{S}_y^{\rm in},
\ee
and if initially
\bea
    \mean{\hat{\bf S}^{\rm in}}=\frac{n_{\! J}}{2}\hat{\bf x},\quad
    (\Delta \hat{S}_x^{\rm in})^2=0,\quad (\Delta \hat{S}_{y,z}^{\rm in})^2
	=\frac{n_{\! J}}{4},
    \label{eqs:Sin}
\eea
where $n_{\! J}$ is the number of photons in the ensemble $\hat{\vecb{J}}$,
then we can effectively replace $\hat{S}_x^{\;\rm in}$ by its macroscopic
expectation value $n_{\! J}/2$. Furthermore, developing the
trigonometric expressions and assuming that
$N_J\chi_J^2\ll 1$, we obtain in leading order
\be
    \hat{S}_y^{\rm out}\approx\frac{n_{\! J}\chi}{2}\hat{J}_z^{\rm in}+\hat{S}_y^{\rm in}.
\ee
Hence a measurement of the $y$ component of the
outgoing light vector gives information about $\mean{\hat{J}_z}$,
while $\hat{J}_z$ itself is not affected by the rotation around
the $z$ axis. 

If such a QND measurement is performed after the 
first beam splitter of the interferometer, then the operator
\be
	\hat{J}'_z\equiv\hat{J}_z^{\rm out}
		-\frac{2}{n_{\! J}\chi_J}\hat{S}_y^{\rm out}
\ee
measures the difference between the $z$ component of
the ensemble's atomic spin vector after the 
second beam splitter and the estimated value after the 
first one. The fluctuations of this operator are 
reduced as compared to $\hat{J}_z^{\rm out}$ \cite{QND},
while the fluctuations of $\hat{J}_y$ are enlarged,
which is depicted in Fig.~\ref{fig:SS}. Hence,
the state of the atomic spin vector is squeezed
in the $z$ direction.


\subsection{III.B. Modified interferometric scheme}

\begin{figure}[h]
\includegraphics[width=0.9\columnwidth]{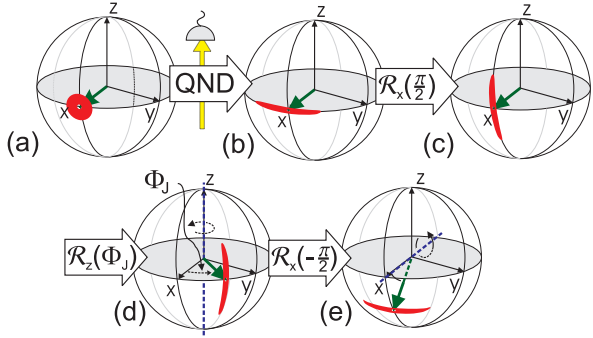}
\caption{(Color online) The interferometric scheme for squeezed input states:
(a) collective spin after the first beam splitter, (b) the 
QND measurement prepares a spin squeezed state which (c) is rotated
around $x$ by $\pi/2$; step (d)$\to$(e) as step (b)$\to$(c) in 
Fig.~\ref{fig:CS}.}
\label{fig:SS}
\end{figure}

We modify the scheme introduced in Section II by 
inserting the QND interaction shortly after the first
beam splitter, followed by an extra rotation
around $x$ by $\pi/2$, which rotates the
uncertainty ellipse such that the phase uncertainty is reduced
as desired, cf. Fig. \ref{fig:SS}.
In the experiment, the latter pulse must not transfer
momentum to the particles, which can be achieved
by using co-propagating Raman lasers for this step,
provided that the two transition frequencies are 
approximately equal, so that the two recoil momenta
cancel in the transitions.

The outgoing spin vector now is calculated as
$\hat{\bf J}^{\rm out}=\rot_x(-\pi/2)\rot_z(\Phi_J)\rot_x(\pi/2)
\rot_z(\chi \hat{S}_z)\hat{\bf J}^{\rm in}$,
and in analogy for the second ensemble $\hat{\bf L}$, with
corresponding Stokes vector $\hat{\bf T}$.

In comparison to Eq.~(\ref{eqs:phiCS}), $\hat{J}_z^{\rm out}$
now has to be corrected to incorporate the spin squeezing (ss)
as described above:
\bea
    \hat{\phi}_{\rm ss}&=&-\frac{1}{N_J}\Big(\hat{J}_z^{\rm out}
        -\frac{2}{n_{\! J}\chi_J}\hat{S}_{y}^{\rm out}\Big)\nonumber\\
    &&+\frac{1}{N_L}\Big(\hat{L}_z^{\rm out}
        -\frac{2}{n_{\! L}\chi_L}\hat{T}_{y}^{\rm out}\Big).
    \label{eqs:phiSS}
\eea
Calculating the expectation value and variance as
before yields
\bea
    \mean{\hat{\phi}_{\rm ss}}&=&\phi
        -\alpha\Big(\frac{1}{N_J}+\frac{1}{N_L}\Big)
		\label{eqn:meansepsq} \\
    (\Delta\hat{\phi}_{\rm ss})^2&=&\frac{1}{n\chi^2}
    \Big(\frac{1}{N_J^2}+\frac{1}{N_L^2}\Big)
        +\alpha\Big(\frac{1}{N_J}+\frac{1}{N_L}\Big).
        \label{eqn:varsepsq}
\eea
Here it has been assumed $n_J=n_L=:n$ and $\chi_J=\chi_L=:\chi$.
Deviations from these assumptions enter the variance only in higher
order terms as long as $||\chi_J|-|\chi_L||/(|\chi_J|+|\chi_L|)\ll1$,
and similar for $n_J$.
Further assumptions leading to these expressions are \mbox{$\bar{N}\chi^2\ll 1$}
as mentioned before, as well as \mbox{
$n\chi^2\ll \sqrt{8}\cdot[\sqrt{\bar{N}}(\theta^2+\phi^2)]^{-1}$}.
Now, provided that $\alpha$ is small enough, the variance is dominated
by the first two terms scaling as $N_{J/L}^{-2}$. They originate from the 
projection noise of the light, and in principle allow
to improve the resolution below the standard quantum limit. However, 
$n\chi^2$ equals, except for a factor of order unity, the fraction of
atoms which are lost due to spontaneous processes during the squeezing 
process, \cf~Section V. Thus,
$n\chi^2\ll1$ is necessary, and even though we obtained a Heisenberg-like
scaling $(\Delta\hat{\phi}_{\rm ss})^2\sim 1/N_{J,L}^2$,
we are far from reaching the Heisenberg limit.

Furthermore, as it becomes clear from the second term, the resolution
is limited by the accuracy of the fluorescence number measurements.
These measurements are necessary in any case to determine the 
phase, because $\mean{\hat{J}_z^{\rm out}}$ and $\mean{\hat{L}_z^{\rm out}}$
have to be rescaled properly. However, $\hat{J}_z$ and $\hat{L}_z$ itself
can be measured using another QND interaction. As will be shown in the next
section, this reduces the dependence on the quality of 
the number measurements.

\subsection{III.C. QND output measurement}
Let us consider now a modification of the scheme using 
squeezed states, where a second QND laser beam is sent
through each ensemble shortly after the last beam splitter. 
We define
\bea
    \hat{\phi}_{\rm ss+}=-\frac{1}{N_J}\frac{2}{n\chi}
    \Big(\hat{S}_{y,r}^{\rm out}-\hat{S}_{y}^{\rm out}\Big)\nonumber\\
    +\frac{1}{N_L}\frac{2}{n\chi}\Big(\hat{T}_{y,r}^{\rm out}
    -\hat{T}_{y}^{\rm out}\Big)
    \label{eqs:phiSSQ},
\eea
where the extra index $+$ is supposed to indicate the 
additional QND measurement. Furthermore, $\hat{\vecb{S}}_r$ and 
$\hat{\vecb{T}}_r$ correspond to individually 
prepared light pulses used to read out $\hat{J}_z$ and $\hat{L}_z$,
respectively.
The resulting expectation value and variance in leading order are
\bea
    \mean{\hat{\phi}_{\rm ss+}}&=&\phi
        +\alpha\theta\Big(\frac{1}{N_J}+\frac{1}{N_L}\Big) \\
    (\Delta\hat{\phi}_{\rm ss+})^2&=&\frac{2}{n\chi^2}
    \Big(\frac{1}{N_J^2}+\frac{1}{N_L^2}\Big)
    \nonumber
    \\
    & & +\alpha(\theta^2+\phi^2)\Big(\frac{1}{4N_J}+\frac{1}{4N_L}\Big).
    \label{eqs:varSSQ}
\eea
Let us compare the modified scheme including an additional 
QND measurement with the scheme using squeezed states.
The magnitude of $\alpha$ in the variance of the modified scheme
is effectively reduced for small $\theta$ and $\phi$,
$\alpha\to\alpha(\theta^2+\phi^2)/4$, 
as compared to Eq.~(\ref{eqn:varsepsq}). Hence, the dependence 
on the number measurement is reduced. 
Furthermore, the leading term shifting the expectation value 
from the desired result $\phi$ is smaller by a factor $\theta$ as 
compared to Eq.~(\ref{eqn:meansepsq}). 

As a further possible advantage, the effect of a non-symmetric 
atom-light interaction is compensated if the coupling of the read-out
QND pulse is similar to the coupling of the squeezing pulse \cite{kuzmichPRL}. 
However, this advantage is probably not very relevant 
for the Sagnac interferometer considered here due to the 
long time-of-flight of the ensembles in the interferometer
between the two QND pulses.

A disadvantage of this scheme is that the first term of the variance 
of Eq.~(\ref{eqs:varSSQ}) comes with
a factor of $2$ because of the two projection measurements of the light
necessary per atomic ensemble. It is possible, although technically demanding,
to reduce this contribution by re-using the light from the first QND 
interaction for the read-out. In this case,
$\chi\to-\chi$ is needed in the second interaction
in order to obtain the difference
$\hat{J}_z^{\rm out}-\hat{J}_z^{\rm in}$ \cite{kuzmichPRL},
and in analogy for the second ensemble.
The  sign can also be achieved by a $\pi$ rotation of the 
atom spin vector around the $x$ axis in between the final beam splitter 
and the QND read-out pulse.


\section{IV. Jointly squeezed ensembles}

In the preceding section we have seen that the $1/(n\chi^2N_J^2)$ term in the
variance comes with a factor given by the number of QND interactions 
with \emph{different} ensembles of light, \cf~Eq.~(\ref{eqs:varSSQ}). 
For this reason let us consider the case of preparing
the initial state of the two atomic ensembles with only
a single QND pulse that interacts with both ensembles consecutively.
The first interaction (with ensemble $\hat{\vecb{J}}$) transforms
$\hat{\vecb{J}}^{\text{in}}\rightarrow\rot_z(\chi \hat{S}_z^{\rm in})\hat{\vecb{J}}^{\text{in}}$,
the second interaction (with ensemble $\hat{\vecb{L}}$) transforms
$\hat{\vecb{L}}^{\text{in}}\rightarrow\rot_z(\chi \hat{S}_z^{\rm in})\hat{\vecb{L}}^{\text{in}}$, 
because $\hat{S}_z^{\rm in}$ itself remains unchanged during the QND interaction. The
Stokes vector $\hat{\vecb{S}}$ transforms as
$\hat{\vecb{S}}^{\text{out}}=\rot_z(\chi \hat{L}_z^{\rm in})
\rot_z(\chi \hat{J}_z^{\rm in})\hat{\vecb{S}}^{\text{in}}$,
and the $y$ component of the outgoing light is given by
\bea
	\hat{S}_y^{\text{out}}&=&\cos(\chi \hat{L}_z^{\rm in})
	\left[\cos(\chi \hat{J}_z^{\,\rm in})\hat{S}_y^{\,\rm in} + \sin(\chi \hat{J}_z^{\,\rm in})\hat{S}_x^{\,\rm in}\right]+\nonumber\\
	&&\,\sin(\chi \hat{L}_z^{\rm in})
	\left[\cos(\chi \hat{J}_z^{\,\rm in})\hat{S}_x^{\,\rm in} - \sin(\hat{J}_z^{\,\rm in}\chi)\hat{S}_y^{\,\rm in}\right],
\eea
such that for $N\chi^2\ll1$, and $\hat{\vecb{S}}$ initially prepared as in
Eq.~(\ref{eqs:Sin}), we have
\be
\hat{S}_y^{\text{out}}\approx\frac{n\chi}2(\hat{J}_z^{\,\rm in}+\hat{L}_z^{\rm in})+\hat{S}_y^{\,\rm in}.
\label{eqn:Syjs}
\ee
Measuring $\hat{S}_y^{\text{out}}$ thus reveals information about $\hat{J}_z^{\rm in}+\hat{L}_z^{\rm in}$
and performs a squeezing operation on this joint operator. Now we
apply the same operations as before to the ensemble $\hat{\vecb{J}}$,
but for the ensemble $\hat{\vecb{L}}$ we perform a rotation by $\pi$
around the $x$ axis before the final measurement. After the extra pulse,
the Sagnac shift $\phi$ is effectively encoded
in the sum of the $z$ components instead of into the
difference.


We define the phase operator for the scheme employing 
jointly squeezed (js) ensembles as
\be 
	\hat{\phi}_{\rm js}=-\Big(\frac1{N_J}\hat{J}_z^{\,\text{out}}
	+\frac1{N_L}\hat{L}_z^{\text{out}}-
	\frac{2}{n\chi\bar{N}}\hat{S}_y^{\,\text{out}}\Big).
\ee
Note that in this definition $\hat{J}_z^{\,\text{out}}$ and 
$\hat{L}_z^{\text{out}}$ are divided by the atom numbers of 
the respective ensembles as before, while the QND measurement 
yields an estimate of $\hat{J}_z^{\,\text{out}}+\hat{L}_z^{\text{out}}$
without such a correction, \cf~Eq.~(\ref{eqn:Syjs}).
%
%
As a consequence we expect that we loose the advantages of squeezing if the
numbers of atoms in the two ensembles differ strongly, {\it i.e.}, if
$\gamma\gg1$. We find in leading order \bea
    \mean{\hat{\phi}_{\rm js}}&=&\phi
        -\frac{2\alpha}{\bar{N}}\\
    (\Delta\hat{\phi}_{\rm js})^2&=&\frac{1}{n\chi^2\bar{N}^2}
        +\frac{2\alpha}{\bar{N}}+
        \frac{\gamma^2}{8\bar{N}^2}.
	\label{eqn:jsvar}
\eea
The importance of the number fluctuations at the preparation stage
is reflected in the fact that in order to arrive at these equations,
the assumption 
\be 
	\gamma\sqrt{\bar{N}}(n\chi^2)^2(\theta^2+\phi^2)\ll 1
	\label{eqn:cond2}
\ee
is necessary in addition to the assumptions leading 
to Eq.~(\ref{eqn:varsepsq}). Furthermore, now $1/(n\chi^2\bar{N}^2)$ 
is the leading term only if $n\chi^2\gamma^2/8\ll1$.

A QND measurement could also be used after the interferometer
to directly read out the joint observable $\hat{J}^{\text{out}}_z+\hat{L}_z^{\text{out}}$
by defining
\be
	\hat{\phi}_{\rm js+}=-\frac{2}{n\chi}\frac{1}{\bar{N}}
	\left(\hat{S}_{y,r}^{\text{out}}-\hat{S}_{y}^{\text{out}}\right),
\ee
where again the indices $r$ refers to the read-out QND measurement.
Notice that in this way it is not possible to measure the
correct rescaled observable \mbox{$\hat{J}_z^{\text{out}}/N_J+\hat{L}_z^{\text{out}}/N_L$},
and consequently there is an important contribution from the difference
of the atom numbers in the expectation value already:
\be
\mean{\hat{\phi}_{\rm js+}}=\phi+
\frac{N_L-N_J}{N_L+N_J}\theta+
\frac{\alpha\phi}{2\bar{N}}.\label{eqn:phijsexp}
\ee
Compared to the variance of the scheme using jointly squeezed ensembles 
without the QND read-out measurement, Eq.~(\ref{eqn:jsvar}),
the dependence of the variance on the number measurements and on the atom 
number difference in the two ensembles is reduced for small $\theta,\,\phi$:
\bea
(\Delta\hat{\phi}_{\rm js+})^2&=&\frac{2}{n\chi^2\bar{N}^2}
        +\frac{\alpha}{2\bar{N}}(\theta^2+\phi^2)+\nonumber\\
        & & \,\,+\frac{\gamma^2}{8\bar{N}^2}\alpha(\theta^2+\phi^2).
\eea
However, the $\gamma$-dependent correction to the
expectation value $\mean{\hat{\phi}_{\rm js+}}$ is only negligible 
compared to the standard deviation
$\Delta\phi_{\rm js+}$ if $\gamma^2\bar{N}n\chi^2\theta^2/8\ll1$. This is
generally a stronger criterion than the limit on $\gamma$ 
encountered without the QND read-out, \cf Eq.~(\ref{eqn:cond2}).

The offset can be compensated by using an estimate for $\theta$ from the
final fluorescence measurement
\be
\hat{\theta}=-\frac1{N_J}\hat{J}_z^{\text{out}}+\frac1{N_L}\hat{L}_z^{\text{out}},
\ee
to define a corrected phase operator
\be
\hat{\phi}^c_{js+}=
\hat{\phi}_{js+}-\frac{N_L-N_J}{N_L+N_J}\hat{\theta}
\ee
which takes into account the bias of $\mean{\hat{\phi}_{\rm js+}}.$
We find that to leading order
\bea
\mean{\hat{\phi}^c_{\rm js+}} &=& \phi +\frac{\alpha\phi}{2\bar{N}}-
\frac{\gamma^2\alpha}{2\bar{N}^2}\\
\left(\Delta\hat{\phi}^c_{\rm js+}\right)^2 &=&
\frac{2}{n\chi^2\bar{N}^2}+\frac{\alpha\phi^2}{2\bar{N}}+\frac{\gamma^2}{8\bar{N}^2}.
\eea
Hence, the $\gamma$-dependent bias in the expectation value is reduced
by a factor $\alpha\gamma/(\bar{N}^{3/2}\theta)\ll1$, while
the $1/\bar{N}$ term in the variance still has a factor $\alpha\phi^2$.

These are the same advantages that we also found in the case of separately
squeezed ensembles with QND read-out, cf.~Eq.~(\ref{eqs:varSSQ}). However,
the contribution of the term proportional to $1/n\chi^2\bar{N}^2$
is reduced by a factor of $2$ in $(\Delta\hat{\phi}^c_{\rm js+})^2$,
because only two instead of four projective measurements of the 
light are necessary in this case.



\section{V. Decoherence}

The attainable squeezing is limited by the absorption
of photons during the interaction between light and atoms \cite{squeezingGaussian}.
Each atom which absorbs and subsequently spontaneously emits a photon
is no longer correlated to the rest of the atoms, but still adds to the
variance. We estimate the number of atoms contributing to such an uncorrelated
background as the number of scattered photons $n\kappa$,
where $\kappa=N_J\chi\Gamma/\Delta$ is the optical density.
Then, in the limit of $\kappa\ll 1$ and for just a single
ensemble, one finds \cite{squeezingGaussian}
\bea
\mean{J_z^{\,\rm out}}&\rightarrow&\mean{J_z^{\,\rm out}}\left(1-\frac{n\chi\Gamma}{\Delta}\right),\label{eqn:dec_ew}\\
\mean{\left(J_z^{\,\rm out}\right)^2}&\rightarrow&\left(1-\frac{n\chi\Gamma}{\Delta}\right)^2\mean{\left(J_z^{\,\rm out}\right)^2}+\frac{n N_J\chi\Gamma}{2\Delta}
\eea
for the collective atomic spin vector and
\bea
\mean{S_z^{\,\rm out}}&\rightarrow&\mean{S_z^{\,\rm out}}\left(1-\frac{N_J\chi\Gamma}{\Delta}\right),\\
\mean{\left(S_z^{\,\rm out}\right)^2}&\rightarrow&\left(1-\frac{N_J\chi\Gamma}{\Delta}\right)^2\mean{\left(S_z^{\,\rm out}\right)^2}+
\frac{N_J^2\chi\Gamma}{4\Delta}
\eea
for the Stokes vector. 
The leading order corrections to the variance $(\Delta\hat{\phi}_{\rm ss})^2$
given in Eq.~(\ref{eqn:varsepsq}) for the case of separately squeezed
ensembles reads
\bea
(\Delta\hat{\phi}_{\rm ss})^2&\rightarrow&
(\Delta\hat{\phi}_{\rm ss})^2+\frac{n\chi\Gamma}{\Delta}\frac1{\bar{N}}+
\frac{2\Gamma}{n^2\chi\Delta},\label{eq:dec}
\eea
and similarly for the other schemes. We will
use this estimate in the following discussion and leave an in-depth analysis
of decoherence processes, e.g., following the lines of \cite{squeezingGaussian},
to further investigations. Generally, according to Eq. (\ref{eqn:dec_ew})
also the expectation value changes due to decoherence. This can be
accounted for by rescaling, and doing so gives only
higher-order corrections to Eq.~(\ref{eq:dec}).

For usual choices of parameters, the last term 
in Eq. (\ref{eq:dec}) is negligible, while the contribution
proportional to $\bar{N}^{-1}$ is comparable in size to the terms
in Eq.~(\ref{eqn:varsepsq}). This limits the coupling $\chi$ and thus
the achievable squeezing. For all other experimental parameters
fixed, there exists an optimal choice for the detuning $\Delta$
and thus for $\chi$ [see Eq. (\ref{eq:chi})]
which minimizes the variance. Taking into account only the first term in
Eq.~(\ref{eqn:varsepsq}), and in the limit $\Delta\gg\Gamma$,
this optimal choice of $\Delta$ leads to a minimal value for
$(\Delta\hat{\phi}_{\rm ss})^2$:
\be
\min_{\Delta}\left[(\Delta\phi_{\rm ss})^2\right]=
\frac{2}{N^{\frac32}\,d}\sqrt{\frac{2\epsilon_0\hbar c A \Gamma}{\omega}}.
\label{eqn:minvardec}
\ee
The scaling is thus no longer as in the Heisenberg limit,
but it is still better than in shot-noise limited measurements
(see also \cite{auzinsh:2004}). 

In addition, the decay of the states $\ket{1}$ and $\ket{2}$ 
during the interferometer step has to be taken into account.
Choosing long-lived hyperfine
ground-state levels to implement $\ket{1}$ and $\ket{2}$ minimizes this decay. Also,
spin squeezed states have been shown to be robust with respect to both, particle
loss and dephasing \cite{stockton:2003}, in contrast to, e.g., GHZ states,
which are maximally fragile under particle losses. 


\section{VI. Comparison of the schemes}

\begin{figure*}[pt]
\includegraphics[width=1\linewidth]{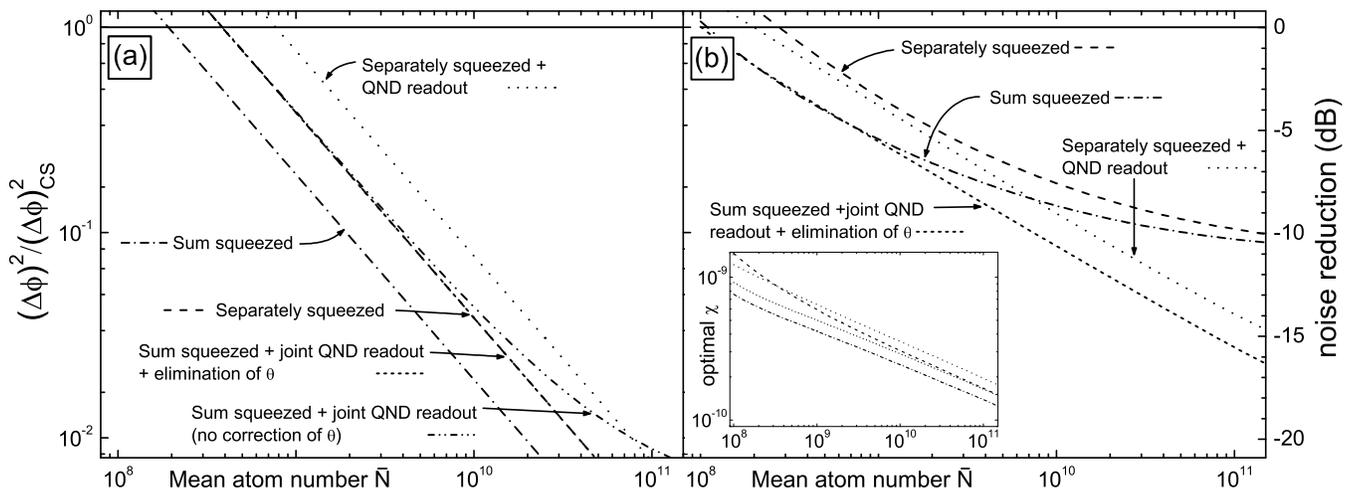}
\caption{
Double-logarithmic plot of the variances
$(\Delta\phi)^2$ for the methods discussed in the text normalized
to the variance for coherent states $(\Delta\phi_{\rm cs})^2$,
as a function of the mean number of atoms $\bar{N}$. The scale on the right
hand site gives the noise reduction in dB.
In both figures $n=10^{11}$, and we fixed $\theta=\phi=0.01$
in order to operate close to the point of maximal sensitivity of
the interferometer. In (a) we consider the close-to-ideal scenario with 
$\alpha=2\times10^{-7}$, $\gamma=10$, and decoherence is not included.
$\chi=3.23\times10^{-10}$ is used, corresponding
to a detuning $\Delta=2.28\times10^{10}\;$s$^{-1}$ for the Rb D$_2$ line
and a cross section $A=0.3\;$mm$^2$ of the laser beam.
In this close-to-ideal scenario, the scaling as $\bar{N}^{-2}$ is visible for all
the methods employing squeezing. The graphs for separately squeezed
ensembles (without QND read-out) and for jointly squeezed ensembles
with QND read-out and corrected expectation value lie on top of each other.
In the case of a joint QND read-out,
failing to correct the expectation value for the contribution from $\theta$ results in a
scaling as $1/\bar{N}$ for $\bar{N}\gtrsim10^9$.
(b) Realistic scenario with $\alpha=2\times10^{-2}$, $\gamma=10^4$,
and including decoherence. In each case and for each value of $\bar{N}$
the detuning has been adjusted to
minimize the variance \cite{parameters}. The inset shows the 
corresponding optimal values of the coupling parameter $\chi$.
For large atom numbers it is clearly seen that the schemes
which do not employ a final QND measurement are strongly
affected by the limitations from the fluorescence detection. The curve for
jointly squeezed ensembles using a joint QND readout lies outside the range of
the figure.
%
}
\label{fig:scaling_all}
\end{figure*}

To analyze the performance of the schemes discussed in the
preceding sections we will fix the number of photons as
$n=10^{11}$ and take $\bar{N}=10^{10}$ as a reasonable
parameter for the mean atom number per ensemble.
We will first consider a close-to-ideal scenario and 
assume that the noise from the fluorescence measurements
can be neglected by setting $\alpha=2\times10^{-7}$. Also, we
will set $\gamma=10$, which for $\bar{N}=10^{10}$ atoms corresponds to
$|N_J-N_L|=10^{-4}\bar{N}$, and we will initially not include decoherence.
Fig.~\ref{fig:scaling_all} (a) shows the scaling with $\bar{N}$ of
$(\Delta\phi)^2/(\Delta\phi_{\rm cs})^2$, {\it i.e.}, of the various variances
normalized to the case of
coherent ensembles. For all the methods involving squeezing of
some observable a Heisenberg like scaling is visible.
The offsets of these curves are given by the
numbers of QND measurements performed, {\it i.e.}, by the factor multiplying 
the $1/(n\chi^2\bar{N}^2)$ term in the variance of each of the considered
schemes that involve squeezing. For the measurement of 
$\phi$ {\em via} reading out $\hat{J}_z+\hat{L}_z$ through
a QND interaction, the term proportional to $\theta$
shifting the expectation value [see Eq.~(\ref{eqn:phijsexp})]
has been included into the variance, in order to
allow for a fair comparison. It is
this term which makes the variance scale only proportional to $1/\bar{N}$
for $\bar{N}\gtrsim10^9$. 
Obviously, correcting this contribution of $\theta$ as described 
in Section IV avoids this term and maintains 
the improvement by a factor of $2$ compared to the
case of squeezing and QND measuring both ensembles separately.
%
%
%
%

In Fig.~\ref{fig:scaling_all} (b), the relative variances are plotted for
realistic experimental parameters and including decoherence. 
$\alpha=2\times10^{-2}$ corresponds to $50$ fluorescence cycles per atom,
and $\gamma=10^4$ is
equivalent to a difference of the number of atoms in the two
ensembles of $10$\% of the mean number $\bar{N}$ at
$\bar{N}=10^{10}$.
With all other parameters fixed, for each value of $\bar{N}$ the interaction
strength $\chi$ is determined by choosing the detuning $\Delta$ from
atomic resonance such that the variance $(\Delta \phi)^2$
is minimized, see Eq.~(\ref{eqn:minvardec}) \cite{parameters}.

As can be seen from Fig.~\ref{fig:scaling_all} (b), in such a realistic 
scenario the noise reduction obtained from
squeezing decreases for all methods. For large atom numbers, in all cases
the variance scales as $1/\bar{N}$ due to decoherence and the noise from the
fluorescence measurements. For all the procedures not involving a QND read-out, 
the strong influence of the latter contribution can be observed, though 
for $\bar{N}=10^{10}$ atoms the total noise still is reduced by
around $7\;$dB with respect to the limit set by quantum projection noise. 
On the other hand, the read-out {\it via} QND measurements allows to reduce the
noise by more than $10\;$dB compared to this limit, while this method does not require an
additional experimental setup as compared to the scheme where QND measurements
are performed only on the incoming atomic ensembles.

While in the close-to-ideal case
squeezing of a joint observable gives an advantage of a factor of $2$ in the variance
compared to individual squeezing and read-out (corresponding to a $3\;$dB noise
reduction), for an experimental reasonable scenario this advantage reduces to 
about $1.5\;$dB. Of course the method of squeezing a joint observable
of both ensembles has a basic advantage since it only needs a single squeezing
operation instead of two, and thus less technical effort is necessary.


\section{VII. Entangled ensembles}\label{sec:entangled}

Julsgaard {\it et al.}~demonstrated experimentally in Ref.~\cite{Polzik}
the generation of macroscopic entanglement between two
atomic ensembles. The scheme to generate such a macroscopically entangled state,
described first in Ref.~\cite{Duan1}, is motivated by
the fact that under the ideal condition of $\gamma=0$
two commuting joint observables can be constructed from $\hat{\vecb{J}}$
and $\hat{\vecb{L}}$:
\begin{equation}
  \mean{[\hat{J}_y-\hat{L}_y,\hat{J}_z+\hat{L}_z]}\propto(N_J-N_L)=0,
  \label{commute}
\end{equation}
{\it i.e.}, $\hat{J}_y-\hat{L}_y$ can be measured without affecting $\hat{J}_z+\hat{L}_z$,
and vice versa. This can be seen directly from
\be
(\rot_z(\chi \hat{S}_z)\hat{\vecb{J}}^{\text{in}}-\rot_z(\chi \hat{S}_z)\hat{\vecb{L}}^{\text{in}})_y=
\hat{J}_y^{\text{in}}-\hat{L}_y^{\text{in}},
\ee
{\it i.e.}, the first QND interaction leaves the difference of
the $y$ components unaffected.
Thus after squeezing the sum $\hat{J}_z+\hat{L}_z$, also the 
difference $\hat{J}_y-\hat{L}_y$ can be squeezed 
without loosing the information gained in the first
measurement. To realize this experimentally in the interferometer,
after the first squeezing
interaction $\hat{\bf J}$ is rotated by a classical $\pi/2$ pulse
around the $x$ axis so that $\hat{J}_y\to \hat{J}_z$ while $\hat{\bf L}$ is
rotated by $-\pi/2$ around $x$ giving $-\hat{L}_y\to \hat{L}_z$. Then a second
laser pulse, prepared again as in Eq. (\ref{eqs:Sin}), interacts consecutively
with both ensembles and thus finally carries information
about $\hat{J}_y-\hat{L}_y$. The outgoing state corresponds to a macroscopically
entangled EPR state \cite{Duan1}.
It is now a natural question to ask whether entangled atomic
ensembles are of use in Sagnac atom interferometry.

For the schemes discussed so far, the collective spin vectors
lie in the $x-z$ plane after the last step of the interferometer.
Therefore always $\mean{\hat{J}_y-\hat{L}_y}=0$,
and an additional operation is necessary in order to
encode phase information in the $y$ components as well.
This can be achieved by rotating both ensemble vectors $\hat{\vecb{J}}$
and $\hat{\vecb{L}}$ by an angle $\varphi$ and $-\varphi$
around the $x$ axis before and after the interferometric phase is applied, respectively.
In this way the plane of rotation of the phase shift is effectively
tilted by $\varphi$ around the $x$ axis.

The measurement process now consists of 
first rotating $\hat{\vecb{J}}$ and $\hat{\vecb{L}}$ by $\pm\pi/2$ and
using another QND interaction to measure the sum of the $z$ components.
This measurement, scaled correctly by a $\varphi$-dependent factor,
reveals $\phi$. To be more explicit, the corresponding operator 
for entangled ensembles (EE) reads
\be
\hat{\phi}_{\rm EE}=\frac1{\cos\varphi}\frac2{\bar{N} n\chi^2}
\left(\hat{S}^{\rm out}_{y,r}-\hat{S}^{\rm out}_{y}\right).
\ee
A measurement of the sum of the $y$ components
can be realized after another rotation around $x$ by either
a QND or a projection measurement. In the former case
\be
\hat{\theta}_{\rm EE}=\frac1{\sin\varphi}\frac2{\bar{N} n\chi^2}
\left(\hat{T}_{y,r}^{\rm out}-\hat{T}^{\rm out}_{y}\right).
\ee
These measurements yield both angles:
\begin{eqnarray}
\mean{\hat{\phi}_{\rm EE}} & = & \phi+
\frac{N_L-N_J}{N_L+N_J}\theta+
\frac{\alpha\;\phi}{2\bar{N}}\\
\mean{\hat{\theta}_{\rm EE}} & = & \theta+
\frac{N_L-N_J}{N_L+N_J}\phi+
\frac{\alpha\;\theta}{2\bar{N}},
\end{eqnarray}
where the offsets can be corrected as above.
For parameters as in Section V. the leading terms
of the corresponding variances read (at $\theta=0$ and $\phi=0$)
\begin{eqnarray}
(\Delta\hat{\phi}_{\rm EE})^2&=&\frac1{\cos^2\varphi}\frac{2}{\bar{N}^2 n\chi^2}+
\frac{\gamma^2\;n\chi^2}{8 \bar{N}}.\label{eqn:sagnac:propop_ent}\\
(\Delta\hat{\theta}_{\rm EE})^2&=&\frac1{\sin^2\varphi}\frac{2}{\bar{N}^2 n\chi^2}+
\frac{\gamma^2\;n\chi^2}{8 \bar{N}}.
\label{eqn:sagnac:trestop_ent}
\eea
Changing $\varphi$ allows to trade in
a lower variance of one component for a higher variance of the other. But
these variances are only scaling with $1/\bar{N}^2$ if $\gamma$ is close to zero,
otherwise the last term $\propto \gamma^2n\chi^2/\bar{N}$ in Eqs.
(\ref{eqn:sagnac:propop_ent}) and (\ref{eqn:sagnac:trestop_ent}) is dominating
the scaling. However, 
$\gamma\approx0$ is an obvious requirement in this case, as
otherwise the commutator does not vanish in Eq. (\ref{commute}), and thus
the two squeezing operations are not compatible.

\section{VIII. Conclusion and outlook}

We have presented and compared in detail several methods to improve the
detection of a differential phase shift of two atomic
interferometers beyond the standard quantum limit, having in mind
especially the application to Sagnac interferometry. For this
purpose, we have analyzed the squeezing of individual and joint
observables and, in both cases, the read-out of the interferometer
{\em via} fluorescence detection of the atoms only or by an additional
QND interaction. 

If decoherence and measurement imperfections
are neglected, all the methods of squeezing improve
the behavior of the variance of 
the differential phase to a $1/\bar{N}^2$ scaling, modified 
by a factor $k/(n\chi^2)\gg1$, 
which is determined by the number $k$ of QND interactions
involved, by the number of photons $n$, and by the coupling $\chi$
between atoms and photons. In the case of jointly squeezed
observables, we found that this limit can only be attained if some
constraints on the difference of the number of atoms in both
ensembles can be fulfilled. In all cases, the achievable
squeezing is limited by decoherence due to the absorption
of photons during the QND measurement.

Using fluorescence measurements to
read out the atomic spins after the interferometer always produces
additional noise scaling as $1/\bar{N}$ due to the photon shot
noise. As an alternative method, a QND measurement can be
employed to read out the final state of the interferometer. Although in
this case fluorescence measurements are still necessary to determine the number
of atoms in the two ensembles, their contribution to
the noise is reduced to a large extent. We have shown that
the best method to achieve this is to perform squeezing and read-out
{\em via} a QND measurement of a joint observable of the two ensembles,
provided that the difference between the number of atoms in the
two ensembles can be made smaller than approximately $10$\% of
the mean number of atoms. This procedure
minimizes the number of QND interactions necessary, thereby minimizing 
the factor multiplying the $1/\bar{N}^2$ term in the variance,
and it reduces the experimental effort. 

Finally, we considered the creation of a macroscopically entangled state
of the two atomic ensembles {\it via} squeezing of two non-local, commuting
observables. We showed that in this case both the sum and the difference
of the phase shifts can be measured with a variance scaling with
$1/\bar{N}^2$, and that  
the relative uncertainty can be shifted between both quantities. However,
this scaling can only be reached here if the number
difference between the two ensembles can be made very small.
Therefore, it would be desirable to identify methods to
control the numbers of atoms in the ensembles, for instance by
employing the superfluid -- Mott insulator transition in an
optical lattice embedded in a weakly confining harmonic potential
\cite{micirac,mibloch}. Controlling
this confinement, which plays the role of a local chemical potential, the number
of atoms in the Mott phase at $T>0$ can be controlled and should only depend
mildly on the total number of atoms in the system. 
A detailed analysis of this idea is left for future investigations.

\section{VIII. Acknowledgments}
It is a pleasure to thank H.A. Bachor, M. Drewsen, J.~Eschner, 
P. Grangier, O.~G{\"u}hne, K.~M{\o}lmer, and A.~Sanpera for 
illuminating discussions.
This work has been supported by the DFG (SPP 1116, SPP 1078, GRK 282, POL 436, CRK, and SFB 407),
by the EC (QUPRODIS, FINAQS), and by the ESF (QUDEDIS).
UVP acknowledges support from the Danish Natural Science Research Council.


\begin{thebibliography}{99}
\bibitem{parameters}
For the parameters in the figure, the optimal values for the detuning
$\Delta$ and the interaction parameter $\chi$ are (at $\bar{N}=10^{10}$ atoms
for the Rb D$_2$ line and a laser beam cross section of $0.3\;$mm$^2$):
separately squeezed ensembles $\Delta=2.45\times10^{10}$s$^{-1}$, $\chi=3.01\times10^{-10}$;
separately squeezed ensembles (QND readout) $\Delta=2.07\times10^{10}$s$^{-1}$, $\chi=3.56\times10^{-10}$;
joint squeezing $\Delta=3.01\times10^{10}$s$^{-1}$, $\chi=2.44\times10^{-10}$;
joint squeezing (QND readout and correction) $\Delta=2.53\times10^{10}$s$^{-1}$, $\chi=2.91\times10^{-10}$.
\end{thebibliography}
\end{document}